\documentclass[prb,amsmath, amssymb, superscriptaddress, aps,twocolumn,letter]{revtex4-1}

\usepackage{graphicx}
\usepackage{dcolumn}
\usepackage{bm}
\usepackage[caption=false]{subfig}

\begin{document}

\preprint{APS/123-QED}

\title{Longitudinal Spin Seebeck effect in Pyrochlore Iridates with Bulk and Interfacial Dzyaloshinskii\textendash Moriya interaction}

\author{Bowen Ma}
\affiliation{
 Department of Physics, The University of Texas at Austin, Austin, Texas 78712, USA\\
}
\author{Benedetta Flebus}
\affiliation{
 Department of Physics, The University of Texas at Austin, Austin, Texas 78712, USA\\
}
\affiliation{Department of Physics and Astronomy, University of California, Los Angeles, California 90095, USA}
\author{Gregory A. Fiete}
\affiliation{
 Department of Physics, The University of Texas at Austin, Austin, Texas 78712, USA\\
}
\affiliation{
 Department of Physics, Northeastern University, Boston, Massachusetts 02115, USA\\
}
\affiliation{Department of Physics, Massachusetts Institute of Technology, Cambridge, Massachusetts 02139, USA}

\date{\today}

\begin{abstract}
The longitudinal spin-Seebeck effect (SSE) in magnetic insulator$|$non-magnetic metal heterostructures has been theoretically studied primarily with the assumption of an isotropic interfacial exchange coupling. Here, we present a general theory of the SSE in the case of an antisymmetric Dzyaloshinskii-Moriya interaction (DMI) at the interface, in addition to the usual Heisenberg form. We numerically evaluate the dependence of the spin current on the temperature and bulk DMI using a pyrochlore iridate as a model insulator with all-in all-out (AIAO) ground state configuration. We also compare the results of different crystalline surfaces arising from different crystalline orientations and conclude that the relative angles between the interfacial moments and Dzyaloshinskii-Moriya vectors play a significant role in the spin transfer. Our work extends the theory of the SSE by including the anisotropic nature of the interfacial Dzyaloshinskii-Moriya exchange interaction in  magnetic insulator$|$non-magnetic metal heterostructures and can suggest possible materials to optimize the interfacial spin transfer in spintronic devices.
\end{abstract}

\maketitle

\section{Introduction}
\label{sec1}
Over the last two decades, the pyrochlore family of iridate compounds, A$_2$Ir$_2$O$_7$, where A is a rare earth element and O is oxygen,  has been subjected to intense scrutiny as the family constitutes a rare class of materials where the energy scales of spin-orbit coupling, Coulomb interaction \cite{wan2011topological,arita2012ab} and the electronic bandwidth are all comparable.\cite{witczak2014correlated,rau2016spin,schaffer2016recent} In particular, the interplay between spin-orbit coupling \cite{qi2011topological,hasan2010colloquium,moore2010birth,ando2013topological} and electron correlations \cite{lee2006doping} makes pyrochlore iridates a promising platform for studying quantum phenomena where topology and magnetic frustration compete on the same footing.\cite{witczak2014correlated,Maciejko:np15} These materials display a metal to insulator transition, accompanied by all-in-all-out spin ordering. \cite{yanagishima2001metal,matsuhira2007metal} Interesting phases of matter such as the axion insulator,\cite{yang2010topological, chen2012magnetic,go2012correlation} fractionalized states,\cite{Maciejko:prl14,Kargarian:prl13,levin2009fractional,ruegg2012topological,kargarian2011competing,maciejko2013topological} and Dirac or Weyl semi-metals,\cite{wan2011topological, ueda2018spontaneous, witczak2012topological, chen2012magnetic, yang2011quantum} emerge with increasing the electron-electron interaction strength. The properties of these phases can be detected in some cases via electrical measurements, as they can lead, e.g., to the anomalous Hall effect. \cite{yang2011quantum}

In the limit of strong electron-electron interaction, pyrochlore iridates behave as a magnetic insulating system with strong spin-orbit coupling. \cite{lee2013magnetic} The Dzyaloshinskii-Moriya interaction (DMI) induced by spin-orbit coupling constitutes the source of non-collinear ground states in these compounds and can give rise to chiral spin textures with nontrivial topology, such as skyrmions. \cite{bogdanov2001chiral}

In contrast to the electrical properties of its metallic and semi-metallic counterparts, a transport theory of the excitations characterizing the insulating phase, i.e., a spin transport theory, remains largely unexplored for systems with non-collinear magnetic orders. \cite{flebus2019interfacial} The nature of the lattice geometry combined with spin-orbit coupling implies there is no conserved component of the spin in the presence of the spin-rotational symmetry breaking DM terms.  However, spin transport can still be well-defined at the interface between these magnetic insulating systems and an adjacent metal.  Motivated by these considerations, here we investigate the interfacial spin transport between a non-collinear magnetic insulator$|$normal metal heterostructure. We model the interfacial interaction between the spin density of the insulating and electronic systems by including both interfacial exchange of the Heisenberg type and DMI interactions.

As a cornerstone for future investigations, we focus on thermally-driven spin transport, i.e., the spin Seebeck effect (SSE), \cite{uchida2008observation,bauer2012spin} at the interface between a pyrochlore iridate and a metal, as shown in Fig.\ref{fig:setup}, where  the heterostructure is directly subjected to a temperature gradient.  A spin current originating in the magnetic insulating system is injected into the metallic system which then coverts the spin current to an electrical current (via spin-orbit coupling) that generates a voltage via the inverse spin Hall effect.\cite{sinova2015spin}

We conduct a systematic study of the dependence of the thermally-driven interfacial spin current on the temperature gradient, the interfacial DMI interaction and the crystalline orientation of the interface. Our results show that the spin current injected into the metal is surprisingly sensitive to the orientation of the interface and the direction of the DM vectors, offering a route for both probing magnetic properties via a spin-transport measurement and engineering efficient heterostructures for the SSE.

Our paper is organized as follows. In Sec.~\ref{sec1}, provide a general background on the physics of spin injection from non-collinear magnetic insulating systems.  In Sec.~\ref{sec2}, we introduce our model for the metallic and the non-collinear magnetic insulating systems, and the exchange and DMI-driven interfacial coupling between their spin densities. Using Kubo formalism, we derive an expression for the spin current flowing across the interface of the magnetic insulator$|$normal metal heterostructure. In Sec.~\ref{sec3}, we introduce a specific model Hamiltonian for the magnetic insulating phase of pyrochlore iridates and determine the ground state and the spectra of the magnetic excitations. Applying our transport theory to these results, we present numerical results for the spin current injected from a pyrochlore iridate into a normal metal. We investigate the dependence of the current on the temperature gradient, the ratio between bulk and interfacial DMI interaction and the crystallographic orientation of the interface. Finally, in Sec.~\ref{sec4}, we discuss our conclusions and possible future directions.

\section{\label{sec2} Model and Approach}
\subsection{\label{sec2A}Lattice Model}
\begin{figure}
\includegraphics[width=0.4\textwidth]{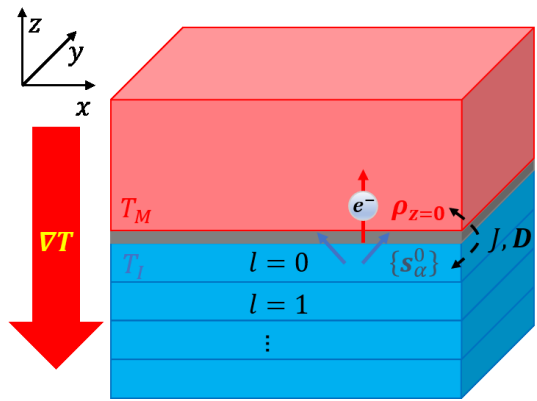}
\caption{\label{fig:setup} (Color online.) Longitudinal Spin Seebeck setup. A thermal gradient $\nabla T$ is applied perpendicular to the interface of the heterostructure. The upper red portion is the non-magnetic metal with temperature $T_M$ near the interface, modeled as a degenerate Fermi gas. The lower blue part is the magnetic insulator with temperature $T_I$, which can be regarded as multiple two dimensional layers with layer index, $l$. The grey part represents the interface, where the electron spin density, $\bm{\rho}_{z=0}$, and $\alpha$-th interfacial magnetic moments, $\bm{S}^0_\alpha$, are coupled with an isotropic Heisenberg exchange $J$ and DM interaction $\bm{D}$.}
\end{figure}

 As illustrated in Fig.\ref{fig:setup}, we consider a magnetic insulator (MI)$|$nonmagnetic metal (NM) heterostructure, allowing the magnetic insulator to have any type of non-collinear magnetic configurations at the interface. A constant temperature gradient is applied across the interface of the heterostructure, yielding a local temperature difference near the interface. A better description may assume a linear temperature variation through the magnetic insulator and into the metallic system, then one would need to express the temperature as a local function of position and express the formulas in terms of quantities with a real-space dependence on the distance from the interface. The physical conclusions of such a calculation would be the same as those we have reached with a simple temperature difference, and very likely the numerical results of a computation would be similar to those we reported in Sec.~\ref{sec3}. For transport problems, it is natural to use periodic boundary conditions in the $xy$-plane and an open boundary at the interface along the transport $z$-direction.

\subsubsection{Nonmagnetic Metal}
Because the non-magnetic material used in experiments is usually a good conducting metal, we treat it as a degenerate Fermi gas with no spin accumulation at the interface. Its spin density at position $(\bm{r},z)$ can be written as,
\begin{align}
\bm{\rho}(\bm{r},z)=\frac{\hbar}{2}\sum_{\sigma \sigma'} \psi^{\dagger}_{\sigma}(\bm{r},z) \bm{\sigma}_{\sigma \sigma'} \psi_{\sigma'}(\bm{r},z),
\label{spindensityelectron}
\end{align}
where $\bm{\sigma}$ is the vector of Pauli matrices with $\sigma$ (or $\sigma'$) the spin index, $\bm{r}$ is a two dimensional in-plane vector, and $ \psi^{\dagger}_{\sigma}(\bm{r},z)$ is the electron creation operator for an electron of spin $\sigma$ at position $(\bm{r},z)$. Because of the translational invariance in the plane parallel to the interface, we can perform a partial Fourier transform with respect to $\bm{r}$, where $ \psi_{\sigma}(\bm{r},z)=\sum_{\bm{k}} \psi_{\bm{k}}(\bm{r}) c_{\bm{k} \sigma}(z)$. The $\bm{k}$ vector is the in-plane wavevector reciprocal to the two-dimensional coordinate $\bm{r}$. The orthonormal basis set $ \psi_{\bm{k}}(\bm{r})$ corresponds to plane waves, i.e.,  $\psi_{\bm{k}}(\bm{r})=e^{i \bm{k}\cdot \bm{r}}/\sqrt{A}$, where $A$ is the area of the interface. The full electronic Hamiltonian can then be written as $\mathcal{H}_{e}= \int\sum_{\bm{k}\sigma}\epsilon_{\bm{k}} c^{\dagger}_{\bm{k}\sigma}(z) c_{\bm{k}\sigma}(z)dz $, with $\langle c^{\dagger}_{\bm{k}'\sigma'}(z') c_{\bm{k}\sigma}(z) \rangle = n_{F}\left( \beta_M \epsilon_{\bm{k}}  \right)  \delta_{\sigma, \sigma'} \delta_{\bm{k},\bm{k}'}\delta_{z,z'}$. Here, $n_{F}(x)=(e^x+1)^{-1}$ is the Fermi distribution function, $\epsilon_{\bm{k}}$ the single-electron energy and $T_M=\beta^{-1}_M$ (we set the Boltzmann's constant $k_B=1$) the local temperature of the metal, which may depend on $\bm{k}$ and z.

\subsubsection{Magnetic Insulator}
Because of the boundary conditions, the magnetic insulator can be regarded as multiple two-dimensional lattices stacked along the $z$-direction.  It is convenient to write the lattice coordinate as $(\bm{R}_{m},l)$, labeling the origin of the $m$-th two-dimensional lattice site in the $l$-th layer. The $m$-th two-dimensional lattice unit cell has a noncollinear magnetic configuration $\{\bm{S}_\alpha(\bm{r}_i)\}$ with a classical macro-spin $\mathbf{S}_{\alpha}(\bm{r}_i)$ located at sublattice $\bm{r}_\alpha$ of lattice $\bm{R}_m$, where $ \bm{r}_{i}=\bm{R}_{m} + \bm{r}_{\alpha}$. For each magnetic site, we can orient the Cartesian coordinate system such that the $\mathbf{\hat{z}}$ axis  locally lies along the classical ground-state orientation of the onsite macro-spin.\cite{laurell2017topological,flebus2019interfacial} Namely, the macrospin $\mathbf{S}_{\alpha}$ is related to the one in the local frame of reference, $\mathbf{S}_{\alpha}'$, as $\mathbf{S}_{\alpha}' = \mathcal{R}_{\alpha} \mathbf{S}_{\alpha}$ with
\begin{eqnarray}
\mathcal{R}_{\alpha}&=&\mathcal{R}_{y}(-\theta_{\alpha}) \mathcal{R}_{z}(-\phi_{\alpha})\nonumber\\
    &=&\left(\begin{array}{ccc}
\cos{\theta_\alpha} & 0 & -\sin{\theta_\alpha}\\
0                   & 1 & 0\\
\sin{\theta_\alpha} & 0 & \cos{\theta_\alpha}
\end{array}\right)\left(\begin{array}{ccc}
\cos{\phi_\alpha}   & \sin{\phi_\alpha} & 0\\
-\sin{\phi_\alpha}  & \cos{\phi_\alpha}   & 0\\
0                   & 0                   & 1
\end{array}\right).\nonumber\\
\end{eqnarray}
Here, the matrix $\mathcal{R}_{z (y)}(\theta)$ describes a right-handed rotation of angle $\theta$ about the $\hat{\mathbf{z}}(\hat{\mathbf{y}})$ axis, and $\theta_{\alpha} (\phi_{\alpha})$ is the polar (azimuthal) angle of the classical ground-state orientation of $\mathbf{S}_{\alpha}$.

In the local frame of reference, the spin at the site $\bm{r}_i$ in the $l$-th layer can be expressed, via the Holstein-Primakoff transformation \cite{holstein1940field} as
\begin{equation}
\left\{\begin{array}{ll}
{S'}_{\alpha,l}^+(\bm{r}_{i})\approx\hbar\sqrt{2S} a_{\alpha}^l(\bm{r}_{i})\\
{S'}_{\alpha,l}^z(\bm{r}_{i})=\hbar\left( S -{a^{l}_{\alpha}}^\dag(\bm{r}_{i}) a_{\alpha}^l(\bm{r}_{i})\right),
\end{array}\right.\label{HP}
\end{equation}
where $S$ is the magnitude of the local spin. Here we ignored higher order terms leading to magnon-magnon interactions. The bosonic operators $a_{\alpha} (a^{\dagger}_{\alpha})$ are related to the  spin-wave annihilation (creation) operators $b_{\beta} (b^{\dagger}_{\beta})$  via a Bogoliubov transformation, \cite{bogoljubov1958new, valatin1958comments} i.e.,
\begin{eqnarray}
a_{\alpha}^l(\bm{q})&=&\frac{1}{\sqrt{N}}\sum_m e^{i\bm{q}\cdot\bm{R}_m}a_{\alpha}^l(\bm{r}_i)\nonumber\\
&=&\sum_{l'}\sum_{\nu=1}^{N_s} \left[ M_{\alpha \nu,\bm{q}}^{ll'} b_{\nu}^{l'}(\bm{q}) +  N_{\alpha \nu,-\bm{q}}^{ll'}{b_{\nu}^{l'}}^{\dagger}(-\bm{q}) \right],\label{BG}
\end{eqnarray}
where  $M_{\alpha \nu,\bm{q}}^{ll'}$ and $N_{\alpha \nu,-\bm{q}}^{ll'}$ are determined from a Bogoliubov transformation that diagonalizes the magnetic excitations of the bulk Hamiltonian of the system, $N$ is the magnetic lattice site number of the two-dimensional layer and $N_s$ is the number of sublattices per unit cell. Thermal magnons in the magnetic insulator follow the Bose-Einstein distribution function $\langle  {\hat{b_{\nu}^l}}^{\dagger}(\bm{k}) \hat{b}_{\nu}^{l}(\bm{k}')  \rangle = n_B\left( \beta_I \omega_{\nu}^l(\bm{k}) \right) \delta_{\bm{k}, \bm{k}'}$, where $n_B(x)=(e^{x} -1)^{-1}$, $\omega_{\nu}^l(\bm{k})$ is the dispersion of the $l$th layer $\nu$th magnonic band, and $T=\beta_I^{-1}$, the local magnon temperature, is below the transition temperature so that the ordering and the fluctuation is well defined.

\subsubsection{\label{sec:level2}Interfacial Coupling}
For simplicity, we assume that the two subsystems have the same Bravais lattice at the interface (corresponding to $z=0$ in the metal or $l=0$ in the insulator), that is, the lattice does not change across the interface. Since the interface generally breaks the inversion symmetry, besides a Heisenberg-type interaction with exchange coupling $J_{ij}$, a Dzyaloshinskii\textendash Moriya interaction (DMI) $\bm{D}_{ij}$ could couple spin density $\bm{\rho}(\bm{r}_i=\bm{R}_m+\bm{r}_\alpha,z=0)$ and magnetic moment $\bm{S}_{\beta,l=0}(\bm{r}_j=\bm{R}_n+\bm{r}_\beta)$, i.e.,
\begin{equation}
    \mathcal{H}_I=\sum_{\langle ij\rangle}\biggl[ J_{ij} \bm{\rho}(\bm{r}_i) \cdot \bm{S}_{\beta}(\bm{r}_j)+\bm{D}_{ij} \cdot \left[\bm{\rho}(\bm{r}_i) \times \bm{S}_{\beta}(\bm{r}_j) \right]\biggr], \label{HI}
\end{equation}
where the summation of $\langle ij\rangle$ is only over the pair of nearest-neighbor sites $i$ and $j$ to avoid double-counts and we drop $z=0$ and $l=0$ in $\bm{\rho}$ and $\bm{S}$, respectively, for notational brevity.

In each local reference frame, by plugging Eq.~(\ref{HP}) and Eq.~(\ref{spindensityelectron}) into Eq.~(\ref{HI}) with Bogoliubov transformation Eq.~(\ref{BG}), we have the interfacial Hamiltonian as
\begin{align}
\mathcal{H}_I&=\frac{\hbar^2\sqrt{2NS}}{2A}\sum_{<\alpha\beta>}\sum_{
\bm{qkk',G}}\sum_{\sigma\sigma'}\sum_{\nu l}\sum_{h}\delta_{\bm{k-k'+q}, \bm{G}}\nonumber\\
&\times\left({g^l_{\alpha\beta\nu}}^{(h)}(\bm{q},\bm{k},\bm{k}'){b_{\nu}^l}^{\dagger}(\bm{q})c^{\dagger}_{\bm{k} \sigma}L^{(h)}_{\sigma\sigma'} c_{\bm{k}'\sigma'} + \text{h.c.}\right),
\end{align}
where $h$ index ($h = 1, 2, 3$) refers the $+, -$ and $z$-components of the spin respectively, $L^{(1, 2, 3)}=\left(\frac{\sigma^+}{2},\frac{\sigma^-}{2},\sigma^z\right)$ and ${g^l_{\alpha\beta\nu}}^{(h)}(\bm{q},\bm{k},\bm{k}')=\left[J_{\alpha\beta}{V^l_{\beta\nu,\bm{q}}}^{(h)}+|\bm{D}_{\alpha\beta}|{U^l_{\alpha\beta\nu,\bm{q}}}^{(h)}\right]e^{-i\bm{(k-k')\cdot r}_{\alpha}}$.

Here, within a mean-field approximation and far from the magnetic ordering temperature (i.e. $\langle{a_{\alpha}^l}^{\dagger}a_{\alpha}^l\rangle \ll S$), we discard the terms $\propto (S-{a_{\alpha}^l}^{\dagger}a_{\alpha}^l)c^{\dagger}_{\bm{k} \sigma}c_{\bm{k}'\sigma'}$ as these are elastic scatterings between electrons off the static magnetic order of the insulator and thus will not depend on the thermal bias. \cite{bender2015interfacial} Then Eq.~(\ref{HI}) can be understood as inelastic scatterings between electrons and thermal magnons in the lowest order, where the contribution from the exchange coupling reads as,
\begin{equation*}
\left\{\begin{array}{lll}
{V^l_{\beta\nu,\bm{q}}}^{(1)}=\left({M_{\beta\nu,\bm{q}}^{0l}}^\dagger\cos^2\frac{\theta_\beta}{2}-N_{\beta\nu,-\bm{q}}^{0l}\sin^2\frac{\theta_\beta}{2}\right)e^{-i\phi_\beta}\\
\\
{V^l_{\beta\nu,\bm{q}}}^{(2)}=\left({N_{\beta\nu,-\bm{q}}^{0l}}\cos^2\frac{\theta_\beta}{2}-{M_{\beta\nu,\bm{q}}^{0l}}^\dagger\sin^2\frac{\theta_\beta}{2}\right)e^{i\phi_\beta}\\
\\
{V^l_{\beta\nu,\bm{q}}}^{(3)}=-\frac{\sin\theta_\beta}{2}\left({M_{\beta\nu,\bm{q}}^{0l}}^\dagger+N_{\beta\nu,-\bm{q}}^{0l}\right),
\end{array}
\right.
\end{equation*}
and the other term from DMI as,
\begin{equation*}
\left\{\begin{array}{llll}
{U^l_{\alpha\beta\nu,\bm{q}}}^{(1)}
=i\left[d^3_{\alpha\beta}{V^l_{\beta\nu,\bm{q}}}^{(1)}-(d^1_{\alpha\beta}-id^2_{\alpha\beta}){V^l_{\beta\nu,\bm{q}}}^{(3)}\right]\\
\\
{U^l_{\alpha\beta\nu,\bm{q}}}^{(2)}
=i\left[(d^1_{\alpha\beta}+id^2_{\alpha\beta}){V^l_{\beta\nu,\bm{q}}}^{(3)}-d^3_{\alpha\beta}{V^l_{\beta\nu,\bm{q}}}^{(2)}\right]\\
\\
{U^l_{\alpha\beta\nu,\bm{q}}}^{(3)}
=i\left[\frac{d^1_{\alpha\beta}-id^2_{\alpha\beta}}{2}{V^l_{\beta\nu,\bm{q}}}^{(2)}-\frac{d^1_{\alpha\beta}+id^2_{\alpha\beta}}{2}{V^l_{\beta\nu,\bm{q}}}^{(1)}\right],
\end{array}
\right.
\end{equation*}
where $\bm{d}_{\alpha\beta}=\frac{\bm{D}_{\alpha\beta}}{|\bm{D}_{\alpha\beta}|}=(d_{\alpha\beta}^1, d_{\alpha\beta}^2, d_{\alpha\beta}^3)$.

The amplitudes of the scattering depends on the rotational angles, $\{(\theta_\beta, \phi_\beta)\}$, and Bogoliubov transformation of the bulk magnetic excitations, $\{({M_{\beta\nu,\bm{q}}^{0l}}, {N_{\beta\nu,-\bm{q}}^{0l}})\}$. Once the geometry of the non-collinear ground state is known, the rotational angles are determined and the latter can be obtained from the magnon Hamiltonian.\cite{del2004quantum, colpa1978diagonalization,valatin1958comments}

\subsection{\label{sec2B}Spin Currents}
To determine the interfacial spin current, we define a total spin accumulation operator $\bm{Q}(z)$ at a $z$-surface as
\begin{align}
\bm{Q}(z)=\int \bm{\rho}(\bm{r},z)\, d\bm{r}=\frac{\hbar}{2}\sum_{\bm{k}}\sum_{\sigma\sigma'}c^{\dagger}_{\bm{k}\sigma}(z) \bm{\sigma}_{\sigma \sigma'} c_{\bm{k}\sigma'}
(z).\label{SurfaceSpin}
\end{align}

Assuming that the magnetic order is static and spin density is conserved across the interface of an area $A$, the interfacial spin current density flowing into the metal can be written as
\begin{align}
    \bm{i}=\frac{1}{A}\frac{d\bm{Q}}{dt}=-\frac{i}{\hbar}\frac{1}{A}[\mathcal{H}_I,\bm{Q}(z=0)]\label{scd}.
\end{align}
By using Kubo formula \cite{kubo1957statistical} to second order in interfacial coupling $J_{ij}$ and $\bm{D}_{ij}$, and after plugging Eq.~(\ref{HI}) and Eq.~(\ref{SurfaceSpin}) into Eq.~(\ref{scd}), some algebra and Wick's theorem \cite{wick1950evaluation} gives,
\begin{widetext}
\begin{equation}
\left\{\begin{array}{lll}
\langle i^x \rangle=\frac{\hbar^3NS}{A^3}\sum_{\nu l}\sum_{hh'} \sum_{\tiny{\substack{<\alpha\beta>\\<\alpha'\beta'>}}}\sum_{\bm{qkk',G}}\Gamma_{\nu l}(\bm{q},\bm{k},\bm{k}')\delta_{\bm{k-k'+q}, \bm{G}}\nonumber\\
\mathfrak{Re}\left[{g^l_{\alpha\beta\nu}}^{(3)*}(\bm{q},\bm{k},\bm{k}'){g^l_{\alpha'\beta'\nu}}^{(2)}(\bm{q},\bm{k},\bm{k}')-{g^l_{\alpha\beta\nu}}^{(3)*}(\bm{q},\bm{k},\bm{k}'){g^l_{\alpha'\beta'\nu}}^{(1)}(\bm{q},\bm{k},\bm{k}')\right]\\
\\
\langle i^y \rangle=\frac{\hbar^3NS}{A^3}\sum_{\nu l}\sum_{hh'} \sum_{\tiny{\substack{<\alpha\beta>\\<\alpha'\beta'>}}}\sum_{\bm{qkk',G}}\Gamma_{\nu l}(\bm{q},\bm{k},\bm{k}')\delta_{\bm{k-k'+q}, \bm{G}}\nonumber\\
\mathfrak{Im}\left[{g^l_{\alpha\beta\nu}}^{(3)*}(\bm{q},\bm{k},\bm{k}'){g^l_{\alpha'\beta'\nu}}^{(1)}(\bm{q},\bm{k},\bm{k}')+{g^l_{\alpha\beta\nu}}^{(3)*}(\bm{q},\bm{k},\bm{k}'){g^l_{\alpha'\beta'\nu}}^{(2)}(\bm{q},\bm{k},\bm{k}')\right]\\
\\
\langle i^z \rangle=\frac{\hbar^3NS}{2A^3}\sum_{\nu l}\sum_{hh'} \sum_{\tiny{\substack{<\alpha\beta>\\<\alpha'\beta'>}}}\sum_{\bm{qkk',G}}\Gamma_{\nu l}(\bm{q},\bm{k},\bm{k}')\delta_{\bm{k-k'+q}, \bm{G}}\nonumber\\
\mathfrak{Re}\left[{g^l_{\alpha\beta\nu}}^{(1)*}(\bm{q},\bm{k},\bm{k}'){g^l_{\alpha'\beta'\nu}}^{(1)}(\bm{q},\bm{k},\bm{k}')-{g^l_{\alpha\beta\nu}}^{(2)*}(\bm{q},\bm{k},\bm{k}'){g^l_{\alpha'\beta'\nu}}^{(2)}(\bm{q},\bm{k},\bm{k}')\right]
\end{array}\right.
\end{equation}
\end{widetext}
where
\begin{eqnarray}
&&\Gamma_{\nu l}(\bm{q},\bm{k},\bm{k}')=\nonumber\\
&&\pi \int \frac{d\omega}{2\pi}\int \frac{d\epsilon}{2\pi} \int \frac{d\epsilon'}{2\pi} \delta (\epsilon - \epsilon' +\omega) A^l_{\nu}(\bm{q},\omega) A(k, \epsilon) A(k', \epsilon') \nonumber \\
&&\Biggl\{ \biggl[1 +n_{B}(\beta_I \omega) \biggr] \biggl[1-n_{F}(\beta_M \epsilon) \biggr] n_{F}(\beta_M \epsilon')\nonumber\\
&&\ \ \ \ \ \ \ \ \ \ \ \ \ \ \ \ -n_{B}(\beta_I \omega) n_{F}(\beta_M \epsilon) \biggl[1-n_{F}(\beta_M \epsilon') \biggr]\Biggr\}.
\label{DB}
\end{eqnarray}
Here, Eq.~(\ref{DB}) reflects the inelastic scattering of $e^-+e^-\rightarrow magnon$  and its reciprocal process in terms of the spectral function for the $\nu$th magnon band at the $l$th layer, $A^l_{\nu}(\bm{q},\omega)$, and the electron  spectral function, $A(\bm{k},\epsilon)$. For a non-interacting clean system, we have $A^l_{\nu}(\bm{q},\omega)=2\pi \delta (\omega-\omega^l_{\nu}(\bm{q}))$ and $A(\bm{k},\epsilon)=2\pi \delta (\epsilon-\epsilon_{\bm{k}})$. Assuming the electronic temperature $T_M$ and the single electron energy $\epsilon_{\bm{k}}$ to be both much smaller than the Fermi energy $\epsilon_{F}$, we treat the electron density of states as a constant $D(\epsilon_{F})=D$. Then Eq.~(\ref{DB}) can be simplified as,
\begin{equation}
\Gamma_{\nu l}(\bm{q})=\pi D^2\omega^l_{\nu,\bm{q}}\left[n_{B}(\beta_M \omega^l_{\nu,\bm{q}}) - n_{B}(\beta_I\omega^l_{\nu,\bm{q}})\right]\label{Gamma}.
\end{equation}
When a thermal gradient drives the system into a non-equilibrium state, a local temperature difference between the two subsystems near the interface breaks the detailed balance and generate a longitudinal spin current. Introducing the notation,
\begin{equation}
    \bm{g}=\left(\frac{g^{(1)}+g^{(2)}}{2}, \frac{i(g^{(1)}-g^{(2)})}{2}, g^{(3)}\right),
\end{equation}
we can express the interfacial spin current density in a more compact way as,
\begin{eqnarray}
\bm{i}&=&i\frac{\hbar^3NS}{A^3}\sum_{\substack{\nu l\\hh'}} \sum_{\substack{<\alpha\beta>\\<\alpha'\beta'>}}\sum_{\bm{qkk',G}}\Gamma_{\nu l}(\bm{q},\bm{k},\bm{k}')\delta_{\bm{k-k'+q}, \bm{G}}\nonumber\\
& &\bm{g}{^l_{\alpha'\beta'\nu}}(\bm{q},\bm{k},\bm{k}')\times\bm{g}{^l_{\alpha\beta\nu}}^*(\bm{q},\bm{k},\bm{k}').\label{SCD}
\end{eqnarray}

One may notice that the spin current depends quadratically on $\bm{g}'s$ so that simply changing the signs of both interfacial couplings $J_{ij}$ and $\bm{D}_{ij}$ will not change the result.

Since rotational invariance of the electron spin at the interface is broken by the magnetic noncollinearity of the insulator, one may notice that, in Eq.~(\ref{HI}), all four kinds of spin configurations $\sigma\sigma'$ of the two-electron scattering can contribute to a magnon creation. Thus, in spin current expressions the product between two amplitudes will have some cross terms as interference, which is zero in the ferromagnetic or collinear antiferromagnetic case. These cross terms are generally non-zero but small compared to the diagonal terms due to the phase summation. In some cases they may even cancel out from the lattice symmetry.

\section{\label{sec3}Numerical Results}
\begin{figure}
\subfloat[\label{PI_unitcell}]{\includegraphics[width=0.24\textwidth]{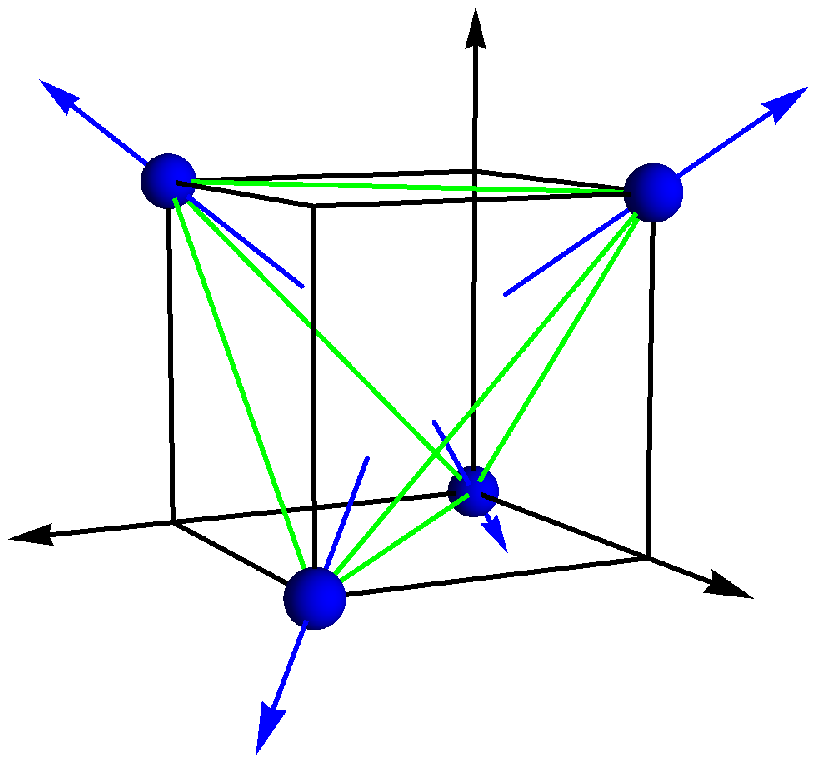}}
\hfill
\subfloat[\label{PI_lattice}]{\includegraphics[width=0.22\textwidth]{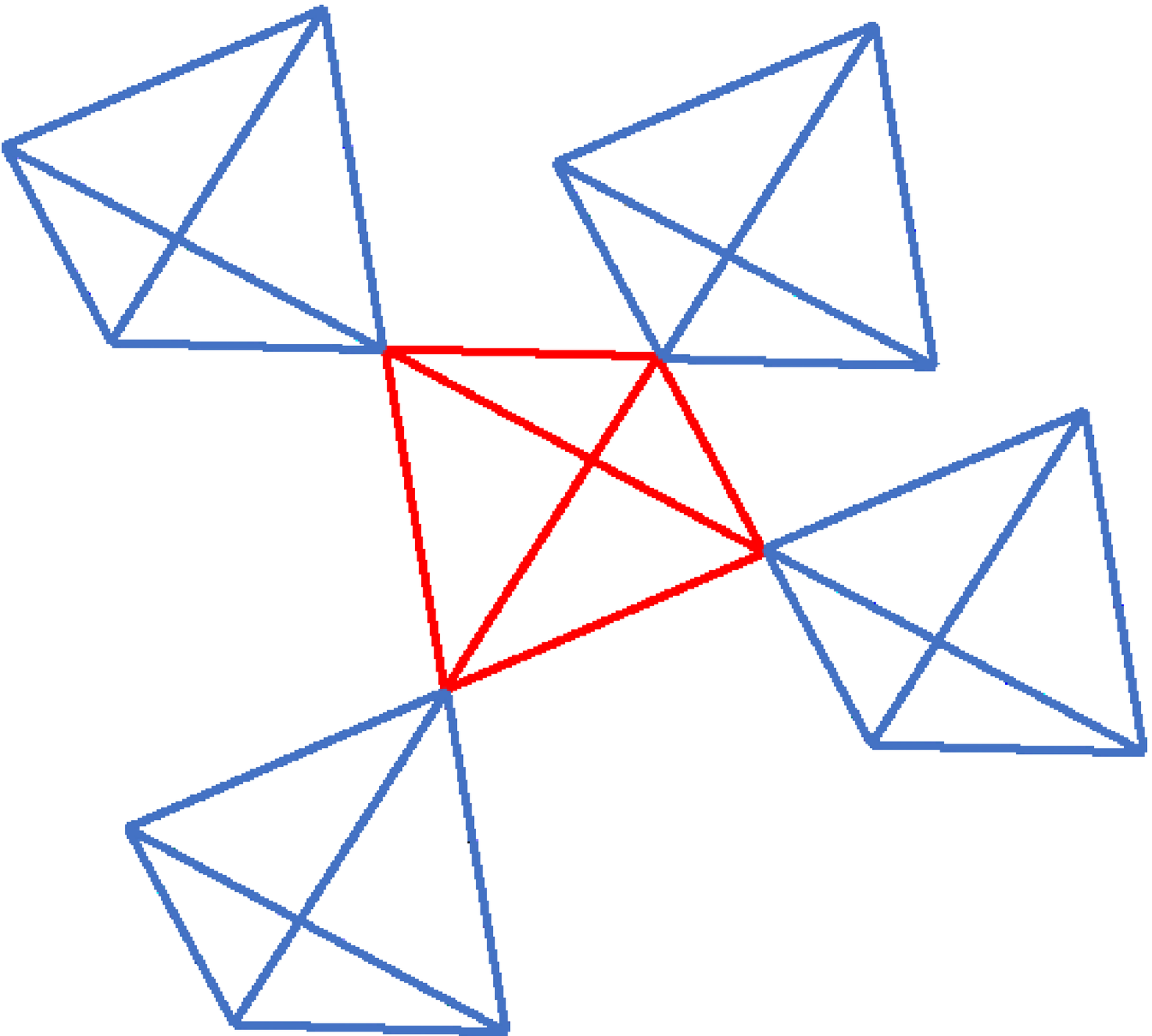}}
\caption{\label{fig:PI_lattice} (Color online.) (a) Bulk unit cell in the cubic coordinate system with the AIAO spin configuration. (b) Corner-sharing pyrochlore lattice.}
\end{figure}

To further study the effects of an interfacial temperature difference, interfacial coupling strength and magnetic non-collinearity on the SSE, we numerically calculate the interfacial spin currents with pyrocholore iridates (PI) as the MI. PIs are 5$d$ transition-metal oxides with a corner-sharing tetrahedron lattice, where on each vertex sits an Ir$^{4+}$ ion, as shown in Fig.\ref{fig:PI_lattice}. It has a strong spin-orbital coupling that may give rise a large DMI both in bulk and at the interface. Along with strong SOC, the large cubic crystal field from oxygen octahedra reduces the energy manifold of Ir$^{4+}$ electrons into a spin-1/2 effective model. \cite{kim2008novel,kim2009phase,kanamori1957theory,zhang2013effective} Under large on-site interaction $U$, the psudo-spins arrange in an all-in/all-out(AIAO) configuration \cite{shinaoka2015phase,disseler2012magnetic,sagayama2013determination,disseler2014direct,donnerer2016all,liang2017orthogonal} and the PI behave as an insulator.\cite{zhao2011magnetic,taira2001magnetic}

In order to use Eq.~(\ref{Gamma}) and Eq.~(\ref{SCD}) to assess the spin current, we need to first evaluate the magnon energy spectrum $\omega_{\nu,\bm{q}}^l$ and the Bogoliubov transformation coefficients $\{({M_{\beta\nu,\bm{q}}^{0l}}, {N_{\beta\nu,-\bm{q}}^{0l}})\}$. These quantities can be obtained by diagonalizing the spin Hamiltonian \cite{witczak2012topological, lee2013magnetic} of the PI,
\begin{equation}
    \mathcal{H}_m=\sum_{\langle ij\rangle} \mathcal{J} S_{i}\cdot S_{j}+\mathcal{D}_{ij}\cdot (S_i\times S_j)+S_i^a\Gamma_{ij}^{ab}S_j^b,\label{H_PI}
\end{equation}
obtained from a large U expansion. Here, $\mathcal{J}$ and $\mathcal{D}_{ij}$ are the exchange interaction and DMI in the bulk, which can be different from interfacial coupling with the metallic system, and $\Gamma_{ij}^{ab}$ is the symmetric anisotropic exchange.\cite{moriya1960anisotropic}

For simplicity, we suppose the metallic side has the same lattice structure as PI and the interfacial coupling has the proper sign (positive $J_{\alpha\beta}$ and indirect\cite{kim2008novel,kim2009phase} $\bm{D}_{\alpha\beta}$) to favor the AIAO configuration at the interface. In a real material system the interfacial magnetic order may differ from that of the bulk, as may either be determined through first principles calculation or experiment. In that case, one should use the interfacial order in Eq.~(\ref{HI}). For the units, we choose energy scale in units of $t^2/U$ using the parametrization studied in Refs.~\onlinecite{lee2013magnetic, laurell2017topological}, and setting $\hbar=1$, and the lattice constant $a=1$.

\begin{figure*}
\subfloat[\label{111_spectrum}]{\includegraphics[width=0.9\textwidth]{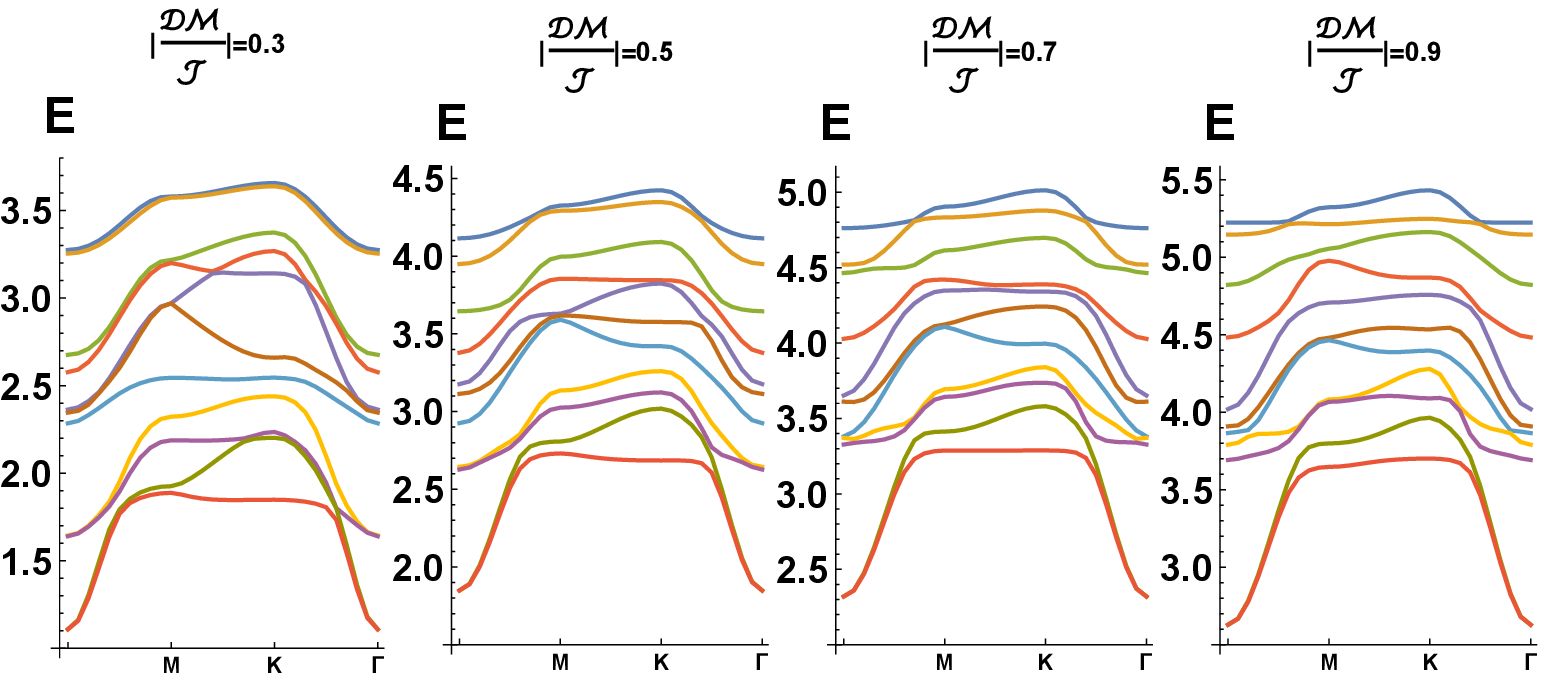}}
\hfill
\subfloat[\label{100_spectrum}]{\includegraphics[width=0.9\textwidth]{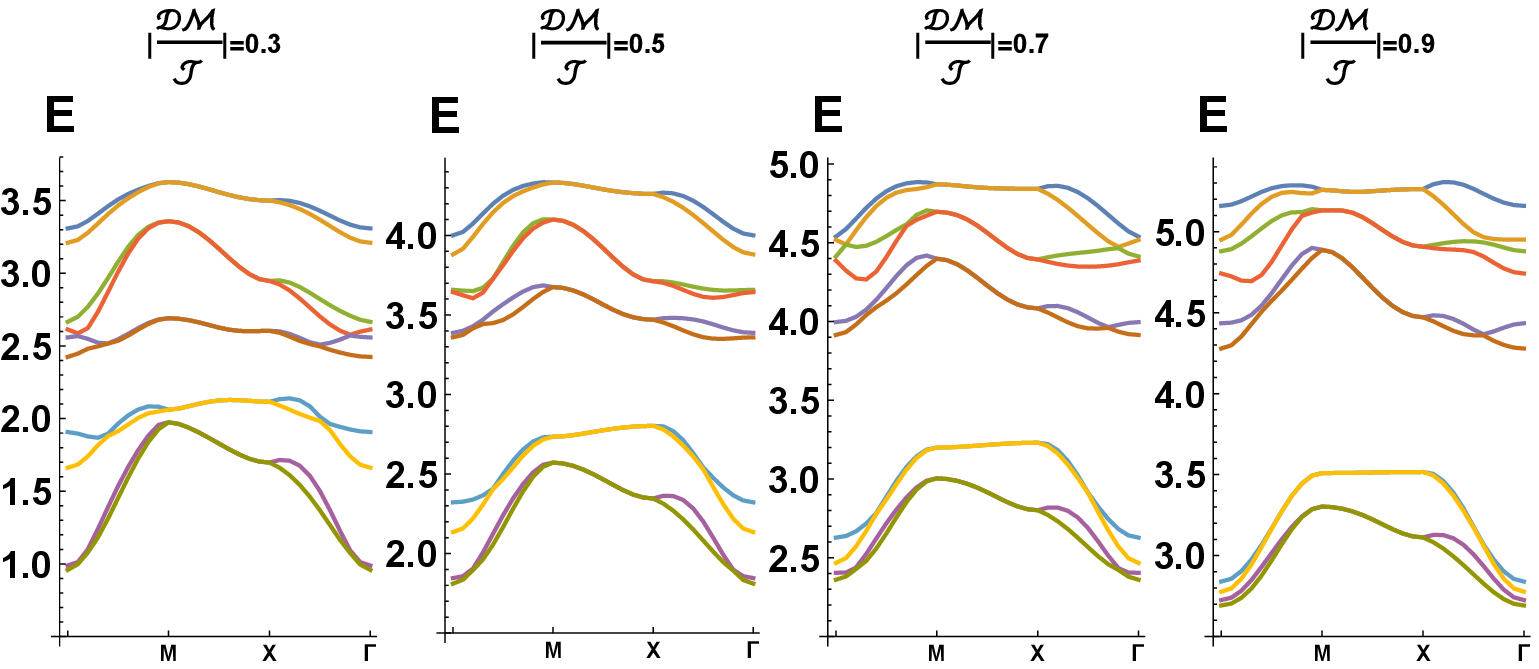}}
\caption{\label{fig:spectrum} (Color online.) Bulk magnon spectrum along high symmetry directions for different crystalline orientations. $\mathcal{|\frac{DM}{J}|}$ is the ratio between the exchange interaction and DMI in the bulk. Energies are given in units of $\frac{t^2}{U}$. (a) Crystalline orientation along [111]. (b) Crystalline orientation along [100].}
\end{figure*}

\begin{figure}
\subfloat[\label{111_config}]{\includegraphics[width=0.26\textwidth]{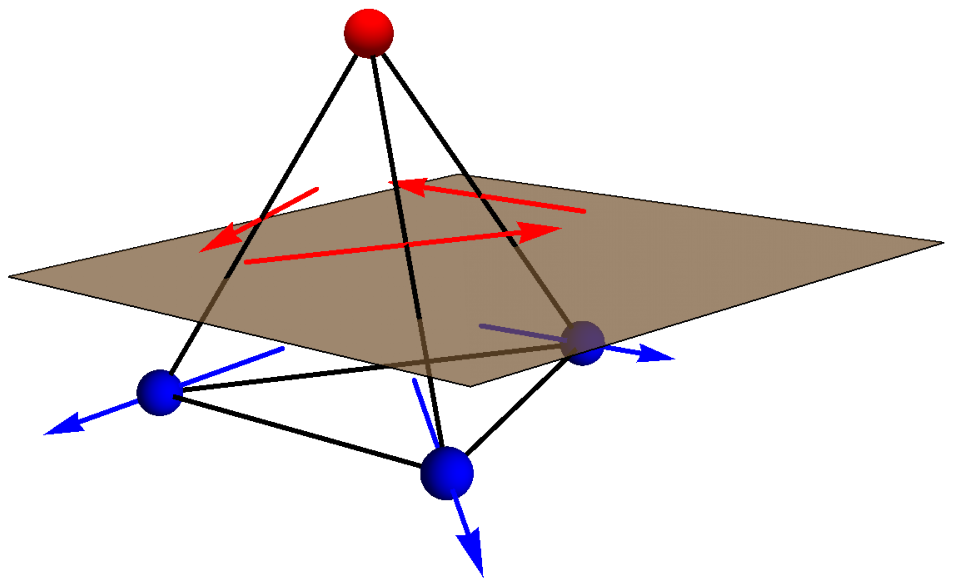}}
\hfill
\subfloat[\label{100_config}]{\includegraphics[width=0.22\textwidth]{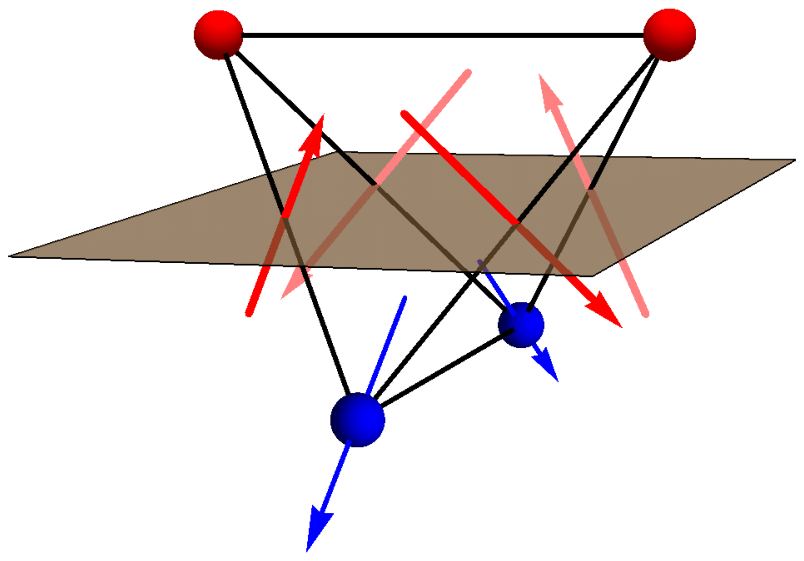}}
\caption{\label{fig:111vs100_config} (Color online.) Blue dots are magnetic sites in the insulator and red dots are spin densities localized at metallic lattices. The brown sheet is the interface and the blue arrows indicate the magnetic moments. The red arrows on the red-blue bonds are the indirect DM vectors $\bm{D}_{ij}$ between electron spin density $\bm{\rho}(\bm{r}_i)$ and moment $\bm{S}_{\beta}(\bm{r}_j)$. Constrained by the symmetry, the DM vector on each bond is parallel to the opposite bond. (a) Crystalline Orientation along [111]. (b) Crystalline Orientation along [100].}
\end{figure}

To analyze the interfacial effect, we consider two separate cases where the nonmagnetic ions of PI are grown in different crystalline orientations, [111] and [100], shown in Fig.\ref{fig:111vs100_config}.
Fig.\ref{fig:111vs100} shows that the spin current flowing through the [111] interface is smaller compared to the [100] case, which can be understood as the interfacial total moment in [100] case is larger than the moment in [111] case.Moreover, we count the contribution both from the scattering that $\bm{G}=0$ and other possible Umklapp scattering ($\bm{G}\neq0$). The system size we use in the numerical summation is a  3-layer$\times12\times12$ lattice to obtain converged results. The dispersion of bulk magnons at the interface calculated from Eq.~(\ref{H_PI}) is shown in Fig. \ref{fig:spectrum}.

\subsubsection{Temperature Dependence}
\begin{figure}
\includegraphics[width=0.47\textwidth]{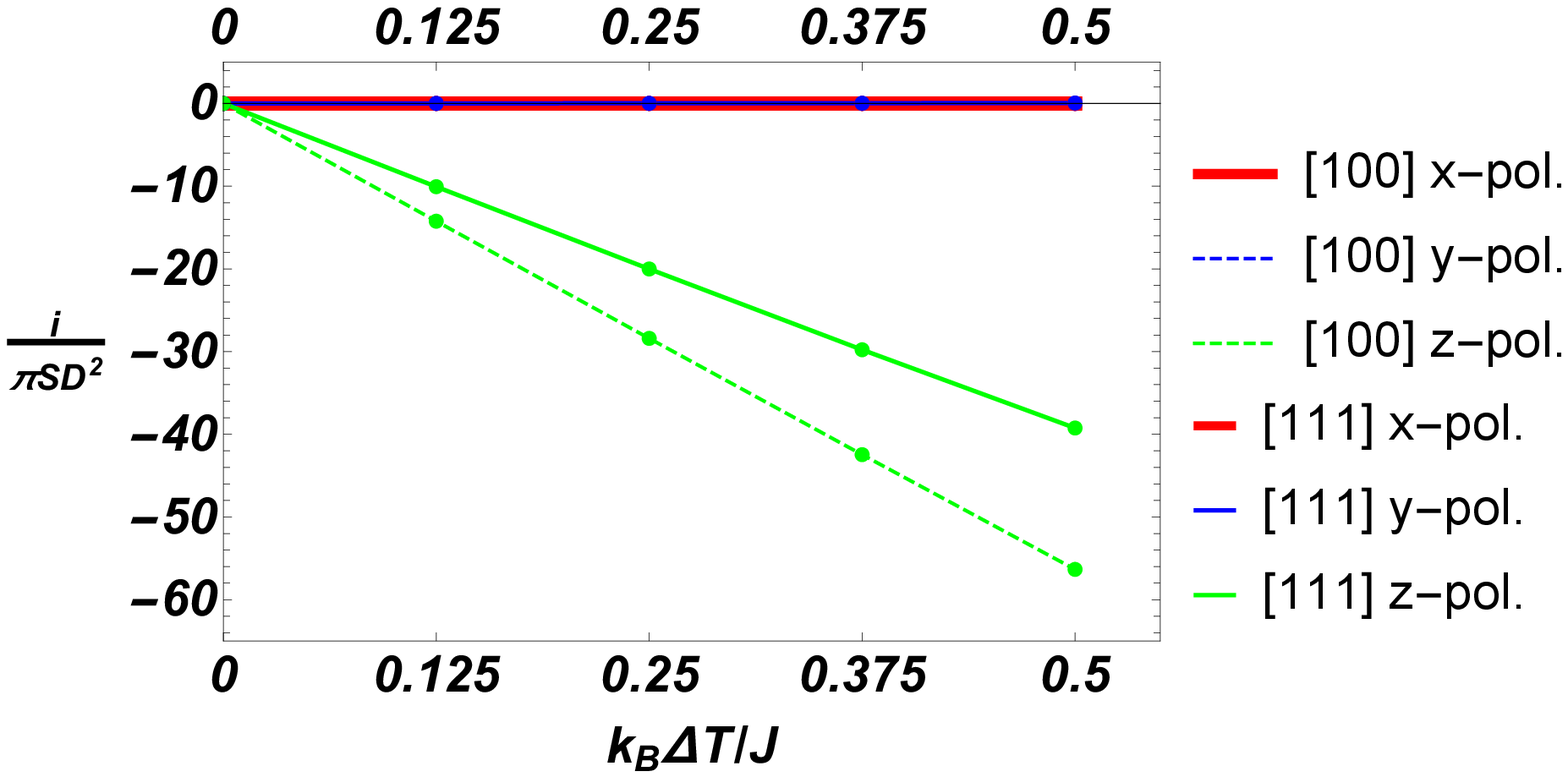}
\caption{\label{fig:111vs100} (Color online.) Red, blue and green lines represent $x$, $y$ and $z$ polarization respectively. Solid (dashed) lines represent spin current density when the crystal is oriented in the [111] ([100]) direction. In these two orientation, only the $z$-polarized spin current density is non-zero (the minus sign means the polarization is along -$z$) and increases as the temperature difference increases, while $i^x$ and $i^y$ is zero because of the symmetry. Here $k_B T_M=\mathcal{J}, k_B T_I=\mathcal{J}-k_B\Delta T$. $|\mathcal{D}_{ij}|/ \mathcal{J}=|\bm{D}_{ij}|/ J_{ij}=0.3$.  $S$ is the magnitude of the local spin and the effective $S$=1/2 in the PI.}
\end{figure}

 We set $k_B T_M=\mathcal{J}$ approximately around the transition temperature of PI and calculate the spin current density when $k_B T_I=\mathcal{J}/2, 5\mathcal{J}/8, 3\mathcal{J}/4, 7\mathcal{J}/8, \mathcal{J}$ in order to investigate the dependence on temperature differences. The nonmagnetic ions of PI are usually grown in the [111] direction; thus, there are 3 sublattices at the interface. As previous experimental studies show that $|\mathcal{D}_{ij}|/ \mathcal{J}\simeq 0.1-0.3$ in the bulk of PI, \cite{donnerer2016all} we set this ratio to be 0.3. Although the strength of the interfacial coupling is not necessarily equal to the bulk one, as the temperatures only show up in Eq.~(\ref{Gamma}) and thus spin currents will not change the dependence on temperature when changing the interfacial coupling, we simply choose interfacial $J_{ij}, \bm{D}_{ij}$ equal to $\mathcal{J}, \mathcal{D}_{ij}$ in the bulk. In both orientations, Fig. \ref{fig:111vs100} shows a linear dependence on temperature difference for $z$-polarized spin current while $x$- and $y$-polarized spin current is almost zero. This can be explained with a symmetry consideration that the net magnetic moment only has a non-zero $z$-component at the interface.

\begin{figure}
\includegraphics[width=0.47\textwidth]{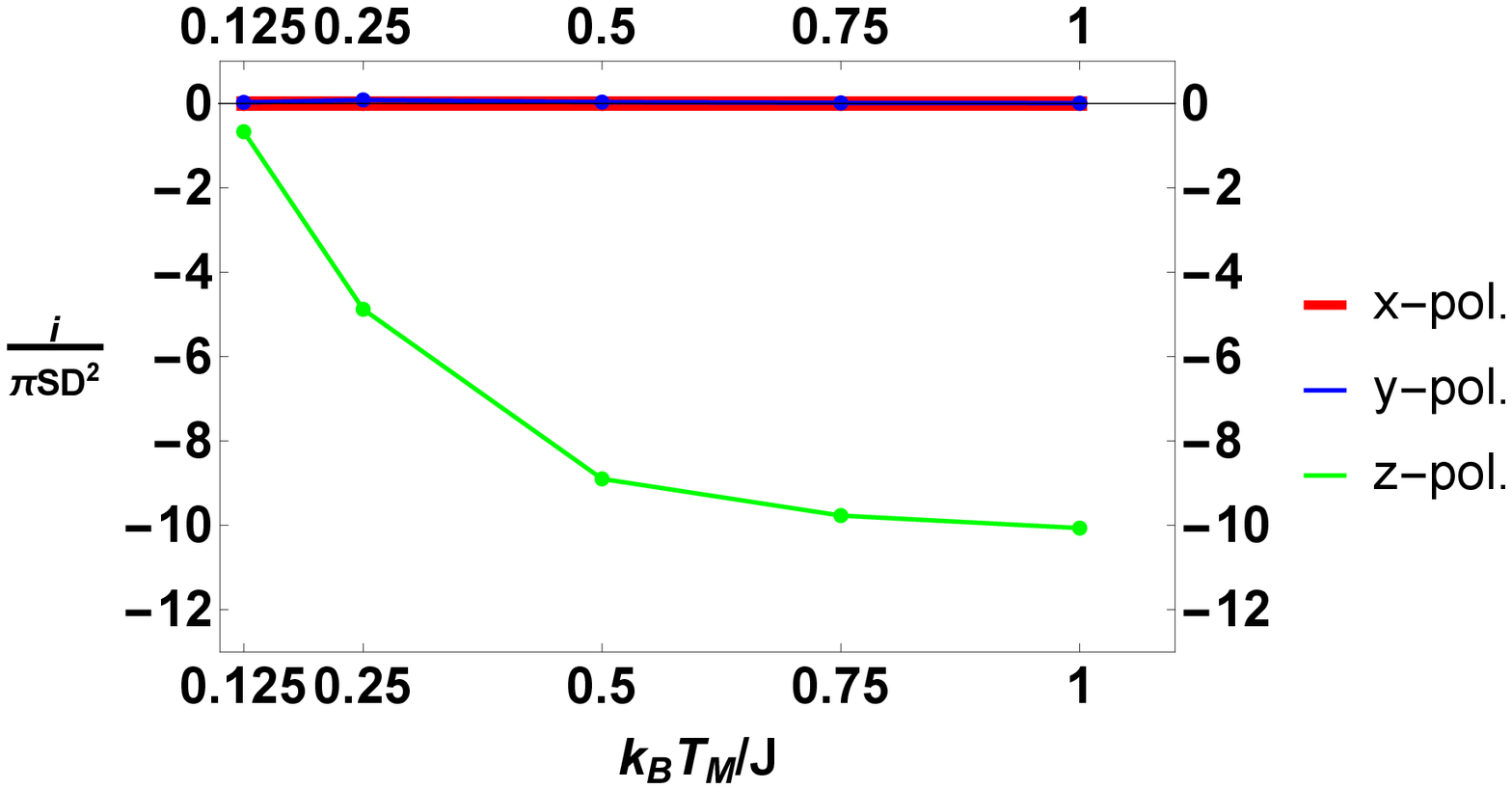}
\caption{\label{fig:base_T}(Color online.) Spin current density increases as the temperature of the heat bath $T_M$ increases. Here $k_B \Delta T=k_B T_M-k_B T_I=\mathcal{J}/8$. $|\mathcal{D}_{ij}|/ \mathcal{J}=|\bm{D}_{ij}|/ J_{ij}=0.3$.}
\end{figure}

Next, we set the temperature difference $k_B\Delta T=\mathcal{J}/8$ as a constant but change the temperature of the heat bath from $k_B T_M=\mathcal{J}$ to $k_B T_M=\mathcal{J}/8$. We find that the spin current density is larger if the whole system is at a higher temperature, as shown in Fig.\ref{fig:base_T}. This comes from the Bose-Einstein statistics of magnons as higher temperature leads to a higher density of magnons participating in the magnon-electron scattering that transfers the spin angular momentum across the interface. However, at high temperature, the magnon-magnon interaction can have a non negligible effects on the scattering which is beyond our present model, and the magnetic order may also changes when the temperature exceeds the transition temperature.

\begin{figure*}
\subfloat[]{\includegraphics[width=0.45\textwidth]{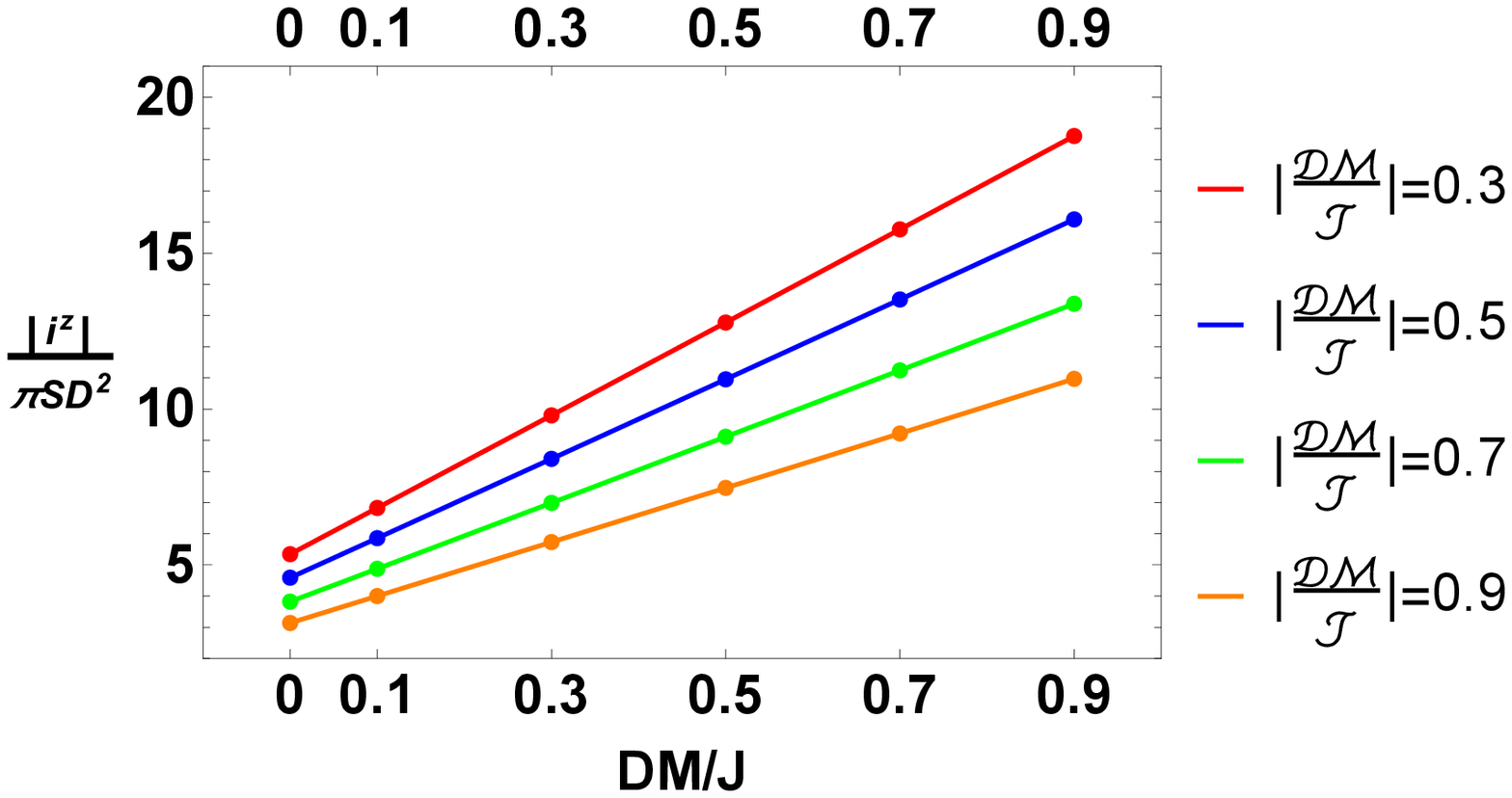}}
\hfill
\subfloat[]{\includegraphics[width=0.45\textwidth]{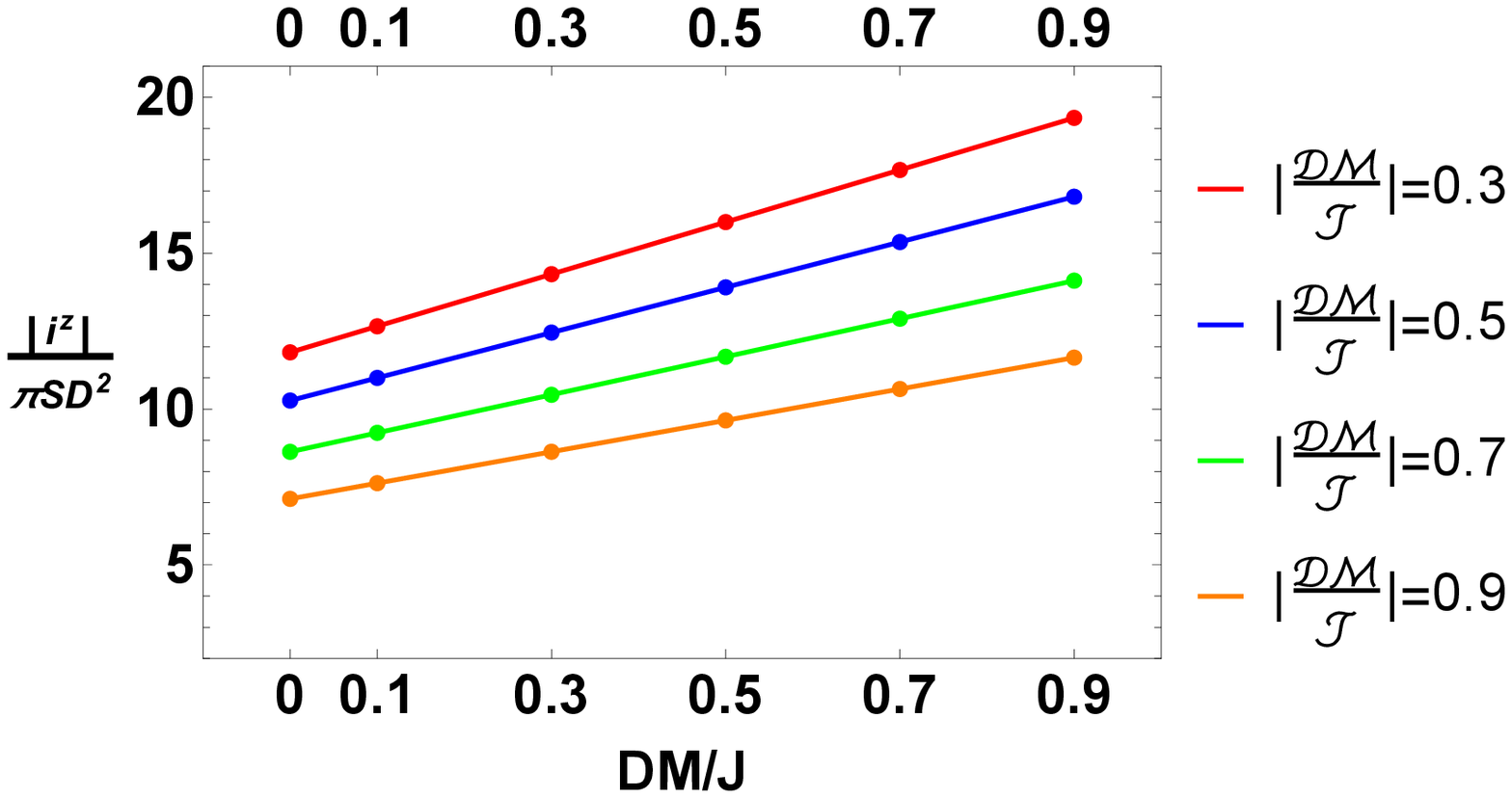}}
\\
\subfloat[]{\includegraphics[width=0.45\textwidth]{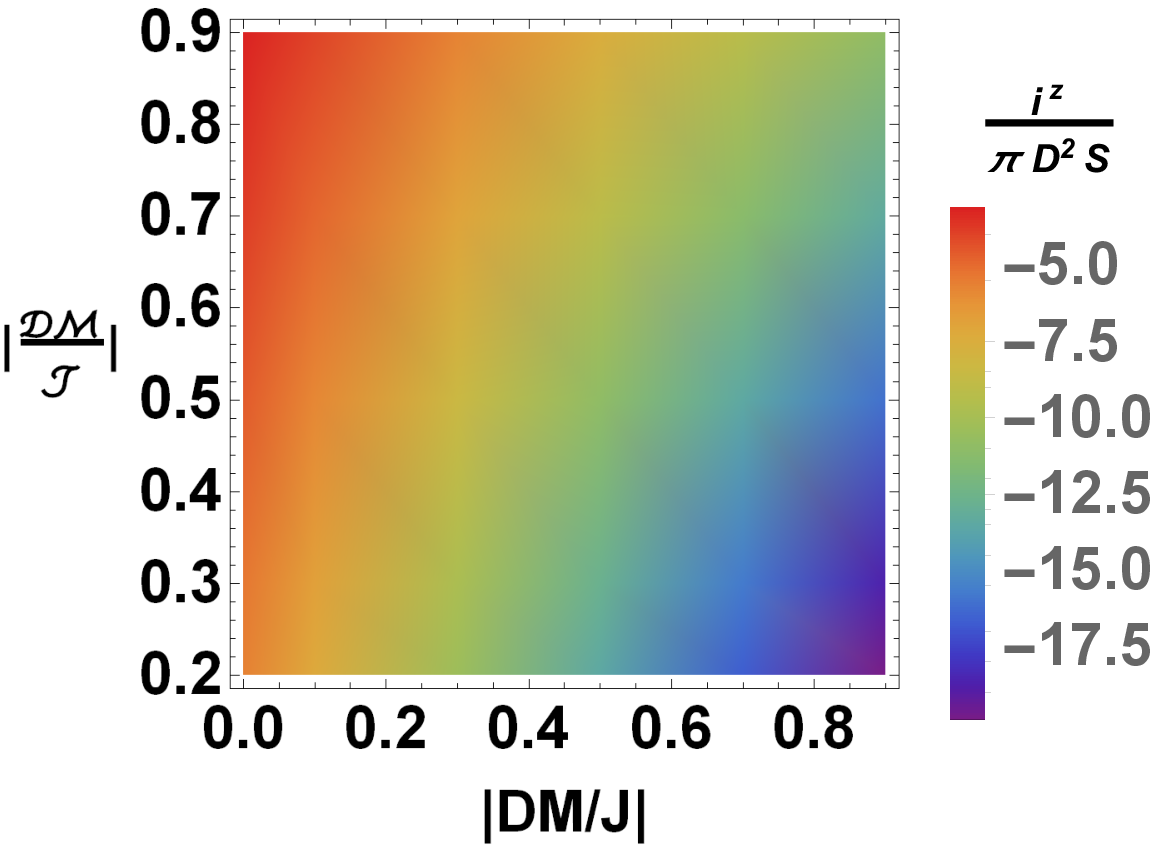}}
\hfill
\subfloat[]{\includegraphics[width=0.45\textwidth]{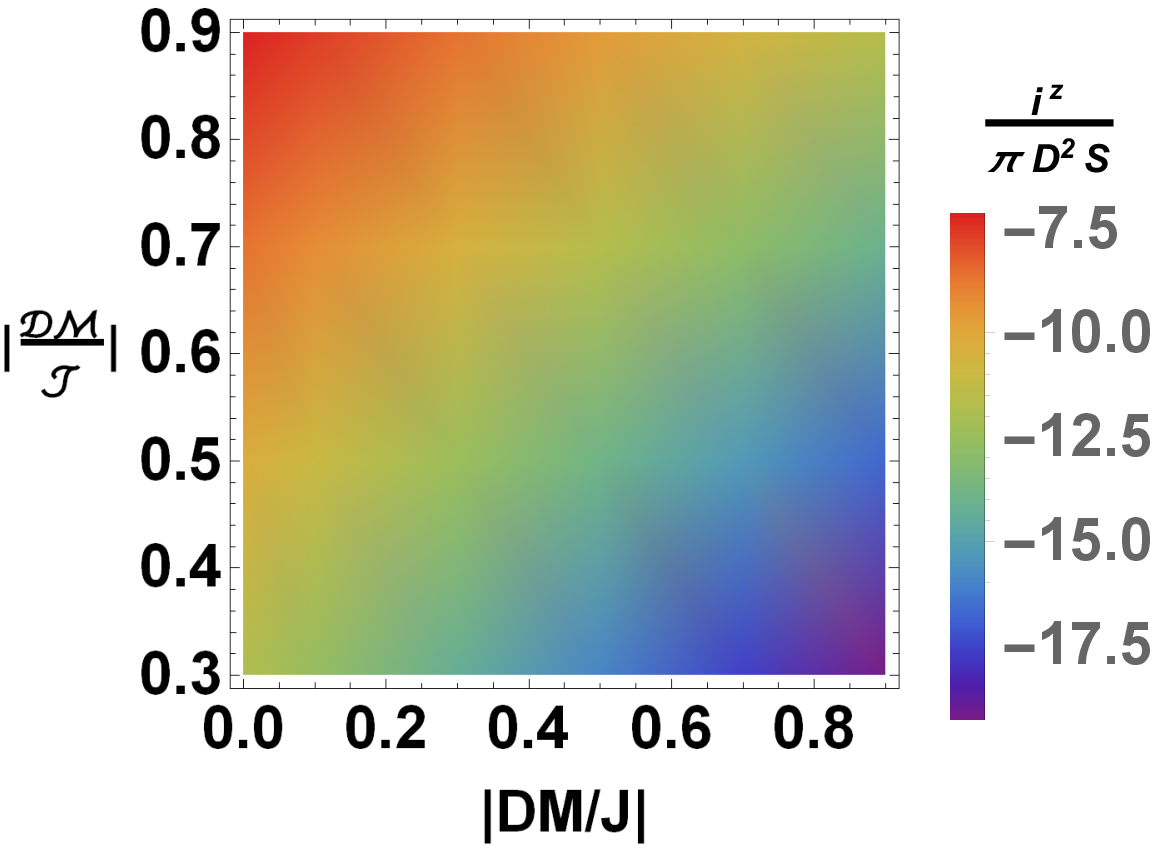}}
\caption{\label{fig:coupling} (Color online.) The magnitude of $z$-polarized spin current density$/\pi D^2S$ as a function of the coupling strength. (a) Crystal oriented in [111] and (b) Crystal oriented in [100]. (c)(d) Corresponding density plots for [111] and [100] orientation. Here $k_B T_M=5\mathcal{J}/8$ and $k_B T_I=\mathcal{J}/2$, $J_{ij}=\mathcal{J}$, $|\frac{\mathcal{DM}}{\mathcal{J}}|$ is the coupling ratio in the bulk and $|\frac{\bm{DM}}{J}|$ is the coupling ratio at the interface. $S$ is the magnitude of the local spin and the effective $S$=1/2 in PI.}
\end{figure*}
\subsubsection{Coupling Strength}
Figure \ref{fig:coupling} shows the effects of the coupling strength on the $z$-polarized spin current when the crystalline orientation is along [111] and [100]. Here, we set $k_B T_M=5\mathcal{J}/8$ and $k_B T_I=\mathcal{J}/2$ but change the ratio of the DMI to exchange coupling $|\frac{\mathcal{D}_{ij}}{\mathcal{J}_{ij}}|$ in the bulk and $|\frac{\bm{D}_{ij}}{J_{ij}}|$ at the interface. As in Eq.~(\ref{SCD}), the interfacial exchange coupling only appears in $\bm{g}$ and $\bm{i}\propto J_{ij}^2$.  Thus, when $|\frac{\bm{D}_{ij}}{J_{ij}}|$ is a constant, the interfacial exchange coupling trivially affects the spin current as a parabolic function of $J_{ij}$. Therefore, we can simply set the $J_{ij}=\mathcal{J}$ and focus on the behavior of spin currents when changing $|\frac{\bm{D}_{ij}}{J_{ij}}|$ and $|\frac{\mathcal{D}_{ij}}{\mathcal{J}_{ij}}|$. Generally, the DMI in the bulk PI increases the excitation energy of magnons, which decreases the spin current. This is shown in Fig.\ref{fig:coupling}: the curve with lower $|\frac{\mathcal{DM}}{\mathcal{J}}|$ is above the one with higher DMI, indicating that the bulk DMI suppresses the spin transport.

One can also see that the spin current has a linear dependence on the interfacial DMI in Fig.\ref{fig:coupling}. To further analyze the effect of interfacial DMI, we consider a FM spin alignment (non-physical, but illustrative) on the pyrochlore lattice, where the magnetic moments are coupled with only exchange interaction $\mathcal{J}<0$ in the bulk and align along transport direction ($z$-direction), while we turn on a DMI at the interface assuming the DM vectors are the same with previous AIAO configuration.

\begin{figure}
\subfloat[]{\includegraphics[width=0.47\textwidth]{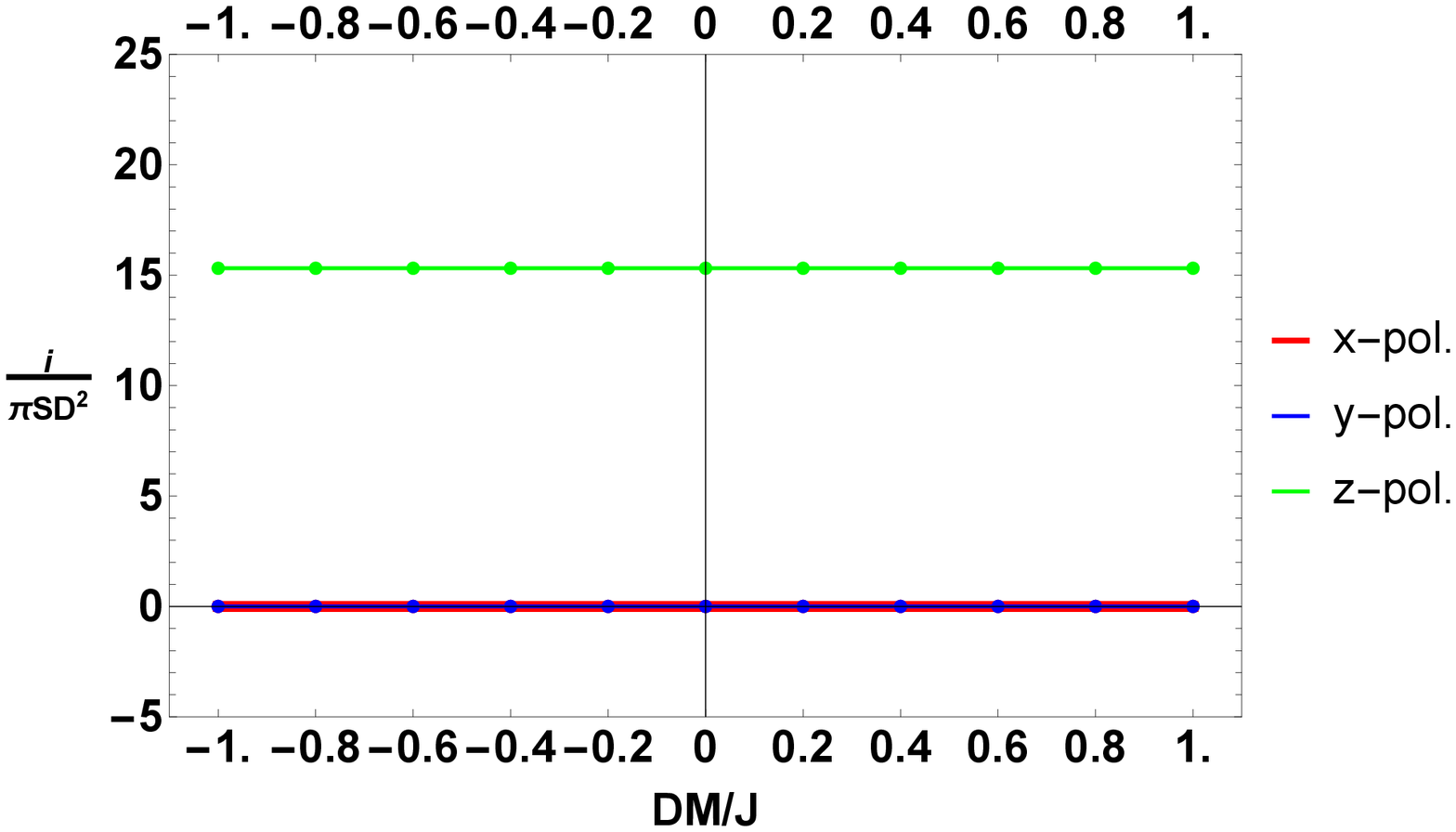}}
\\
\subfloat[]{\includegraphics[width=0.47\textwidth]{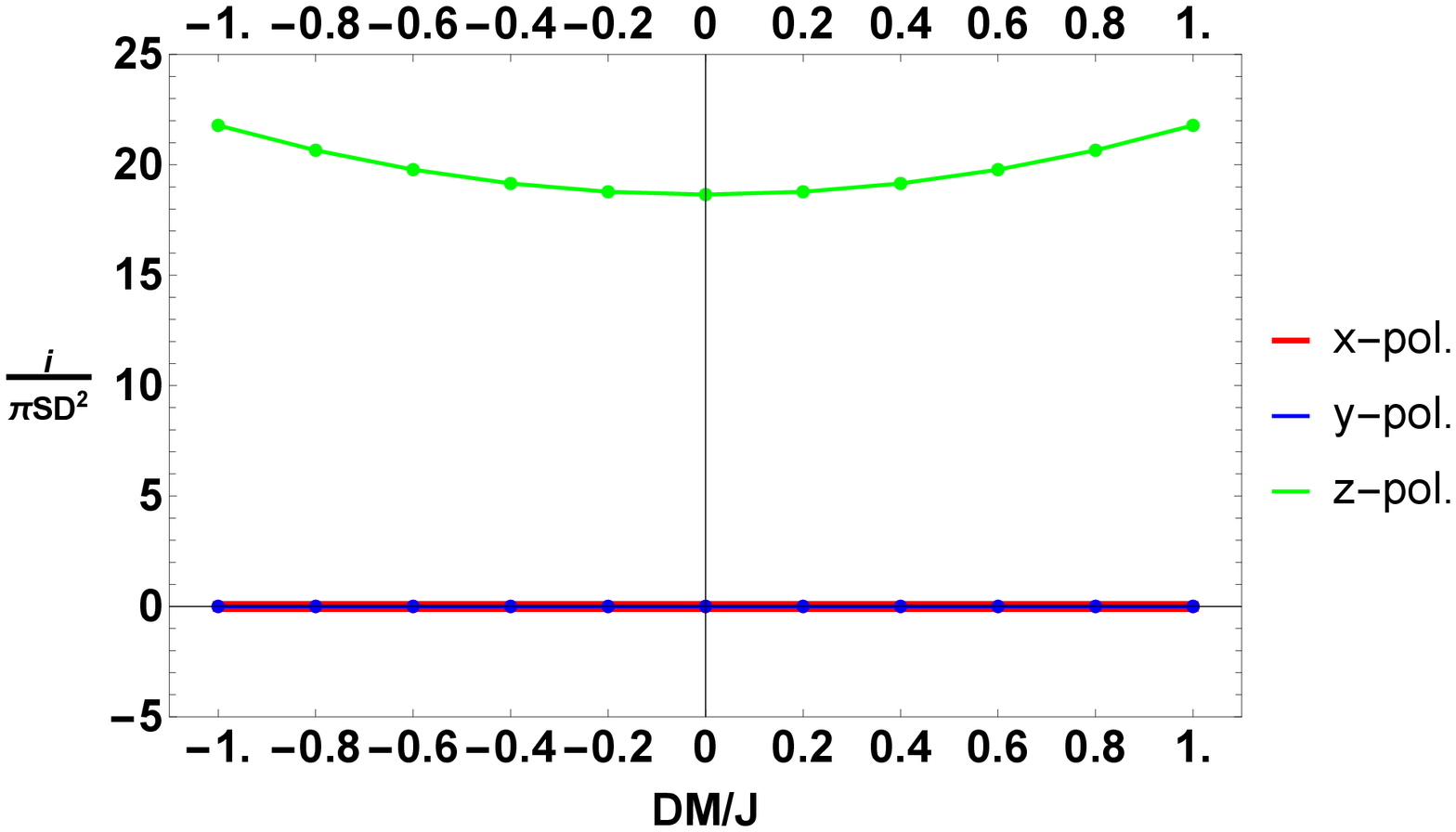}}
\caption{\label{fig:ferro} (Color online.) Spin current density induced by ferromagnetic moments on PI lattice. (a) Crystalline Orientation along [111]. (b) Crystalline Orientation along [100]. In these two orientations, the configurations of interfacial DM vectors are the same with the DM vectors in the AIAO state and we allow the interfacial exchange coupling to be either ferromagnetic ($J<0$) or antiferromagnetic ($J>0$).}
\end{figure}

As shown in Fig.\ref{fig:ferro}, only the $z$-polarized spin currents are non-zero in both cases but the dependence on the interfacial DMI are unlike each other and  unlike the curves in Fig.\ref{fig:coupling}. This difference comes from the relative angles among the magnetic moments, the DM vectors, and the polarization direction. Since the interference is small, if we ignore the cross terms in Eq.~(\ref{SCD}), the spin current density can be simplified as,
\begin{eqnarray}
&\bm{i}&=\frac{\hbar^3N}{A^3}\sum_{\bm{qk,G}}\sum_{\alpha\beta}\sum_{\nu l}\Gamma_{\nu l}(\bm{q},\bm{k},\bm{k}+\bm{q}-\bm{G})\Lambda^l_{\beta\nu}(\bm{q})\nonumber\\
& &\left[\bm{D}_{\alpha\beta}\left(\bm{S}_\beta\cdot\bm{D}_{\alpha\beta}\right)+J_{\alpha\beta}\left(\bm{D}_{\alpha\beta}\times\bm{S}_\beta\right)+J_{\alpha\beta}^2\bm{S}_\beta\right],\label{SCD_simp}
\end{eqnarray}
where $\Lambda^l_{\beta\nu}(\bm{q})=\frac{1}{2}\left(|M_{\beta\nu,\bm{q}}^{0l}|^2-|N_{\beta\nu,-\bm{q}}^{0l}|^2\right)$.

As it can be seen from Eq.~(\ref{SCD_simp}), the spin current density is generally parabolic with respect to the interfacial DMI. The third term within the bracket is the contribution from the isotropic coupling and is quadratic in the exchange coupling strength,\cite{flebus2019interfacial} while the first two terms arise from the interfacial DMI. Since the DMI can be understood as the super-exchange interaction with the help of spin-orbital coupling in a microscopic picture,\cite{moriya1960anisotropic} the total spin is not conserved at the interface. The new contributions come from the orbital moment of the ion. In the case of the AIAO spin configuration, the magnetic moment $\bm{S}_\beta$ is perpendicular to the corresponding DM vectors leading the first term to be zero and giving rise a linear dependence. Moreover, the indirect DM vectors in the AIAO state turn out to lead to an enhancement of spin current in the second term, as shown in Fig.\ref{fig:coupling}.

In the ferromagnetic cases, when the crystal is oriented in [111], both $\bm{S}_\beta\cdot\bm{D}_{\alpha\beta}=0$ and $\left[\bm{S}_\beta\times\bm{D}_{\alpha\beta}\right]_z=0$, which results in the DMI not affecting the spin current in Fig.\ref{fig:ferro}(a). However, $\bm{S}_\beta$ is not perpendicular to $\bm{D}_{\alpha\beta}$ anymore if the orientation is in [100] and this will give a parabolic curve, as seen in Fig.\ref{fig:ferro}(b). More specifically,  $\sum_{\alpha\beta}\Lambda_{\beta\nu}^l\left[\bm{D}_{\alpha\beta}\left(\bm{S}_\beta\cdot\bm{D}_{\alpha\beta}\right)\right]_z=\sum_{\alpha\beta}\frac{1}{2}|M_{\beta\nu,\bm{q}}^{0l}|^2 S_\beta {D_{\alpha\beta}^z}^2$ is always postive and $\sum_{\alpha\beta}\Lambda_{\beta\nu}^l J_{\alpha\beta}\left[\bm{D}_{\alpha\beta}\times\bm{S}_\beta\right]_z=JS\Lambda_{1(2)\nu}^l\sum_{\alpha}\left[\bm{D}_{\alpha,1}\times\hat{z}+\bm{D}_{\alpha,2}\times\hat{z}\right]_z=0$ because of the mirror symmetry of the lattice (where $\Lambda_{1\nu}^l=\Lambda_{2\nu}^l$ and $\bm{D}_{\alpha,1}+\bm{D}_{\alpha,1}\perp \hat{z}$). This leads to a parabola opening up and centered at zero as shown in Fig.\ref{fig:ferro}(b).
Therefore, not only the spin orientation at the interface but also the directions of interfacial DM vectors can affect the interfacial spin current. Whether the interfacial DMI will enhance or suppress the spin transfer depends on the details of interfacial orbital moments and Eq.~(\ref{SCD_simp}) can be used to theoretically investigate the effects.

\section{\label{sec4}Discussion and conclusions}
In this work, we extend the theory of spin transport driven by thermal gradient at the interface of a noncollinear magnetic insulator$|$normal metal heterostructures, \cite{flebus2019interfacial} deriving a general expression for the spin current density when both exchange coupling and Dzyaloshinskii\textendash Moriya interaction are present at the interface. Our theory, which neglects magnon-magnon interactions, is valid at temperatures much below the magnetic ordering temperatures. We numerically calculate the spin current density to study the effects of temperature difference, coupling strength and crystalline orientation using pyrocholore iridates with the AIAO state as the magnetic insulator. We derive an approximate equation to investigate the effects of DM vectors on spin currents and use it to explain the numerical results when the PI is in the AIAO state and the ferromagnetic state. These results give theoretical guidance to manufacturing practical spintronic devices and optimizing the spin transfer across the interface.

Experimentally, longitudinal SSE is detected by the transverse voltage induced from the inverse spin Hall effects. \cite{saitoh2006conversion, valenzuela2006direct,kimura2007room, uchida2010observation} Our result of interfacial spin currents can then be used as a boundary condition to solve the transport equation for spin diffusion and find the spin currents in the bulk of the metal.\cite{sinova2015spin} Making use of experimental data, one can theoretically evaluate the spin Hall angle and investigate different microscopic models on spin-current conversion. This could provide a better description of the relation between the bulk spin Hall angle and the interface properties which may influence the value of the spin Hall angle extracted in inverse spin Hall measurements.

While here we focus on the spin transport due to magnon-electron scattering, thermalized phonons could also contribute to the scattering. \cite{adachi2010gigantic,adachi2011linear,flebus2017magnon,kikkawa2016magnon} Future work should evaluate the correction to the phonon-dressed magnons. As the SSE is a non-equilibrium phenomenon, the interplay between the higher order magnons and the spin accumulation near the metallic interface should be considered. Moreover, in Sec.~\ref{sec3}, we simply treat the interfacial spin configuration and DM vectors as being the same as those in the bulk of the material. However, the lattices of the two systems may not match with each other as we have assumed and the physics at the interface may be more complicated. A first-principle calculation on magnetic canting and orbital moments can be useful to lead to a more realistic result, still within the framework of our theory. In addtion, the asymmetry of the interface can give rise a Rashba-type coupling and Dresselhaus SOC which may introduce Skyrmions near the interface. \cite{fert2017magnetic,banerjee2014enhanced, rowland2016skyrmions}.  These effects should be investigated in future work.

\begin{acknowledgments}
The authors would like to thank P. Laurell and Y. Tserkovnyak for helpful discussions. This work was supported by NSF Grant No. DMR-1949701,  NSF Materials Research Science and Engineering Center Grant No. DMR-1720595, the Simons Foundation, and a QuantEmX grant GBMF5305 from ICAM and the Gordon and Betty Moore Foundation.
\end{acknowledgments}

\nocite{*}

\bibliography{apssamp}

\end{document}